\documentstyle[psfig]{mn2e}

\begin{document}

\title{Constraining dark energy from the abundance of weak gravitational lenses}
\author[N. N. Weinberg and M. Kamionkowski]{Nevin N. Weinberg and
Marc Kamionkowski \\
California Institute of Technology, Mail Code 130-33, Pasadena, CA 91125 USA}


\pagerange{\pageref{firstpage}-\pageref{lastpage}}
\date{September 2002}
\pubyear{2002}

\maketitle

\label{firstpage}

\begin{abstract}
We examine the prospect of using the observed abundance of weak gravitational 
lenses to constrain the equation-of-state parameter $w = p/ \rho$ of the dark
energy. Dark energy modifies the distance-redshift relation, the amplitude of 
the matter power spectrum, and the rate of structure growth. As a result, it 
affects the efficiency with which dark-matter concentrations produce detectable 
weak-lensing signals. Here we solve the spherical-collapse model with dark 
energy, clarifying some ambiguities found in the literature. We also provide 
fitting formulas for the non-linear overdensity at virialization and the 
linear-theory overdensity at collapse. We then compute the variation in the 
predicted weak-lens abundance with $w$. We find that the predicted redshift 
distribution and number count of weak lenses are highly degenerate in $w$ and 
the present matter density $\Omega_0$. If we fix $\Omega_0$ the number count 
of weak lenses for $w = -2/3$ is a factor of $\sim 2$ smaller than for the 
$\Lambda$CDM model $w = -1$. However, if we allow $\Omega_0$ to vary with $w$ 
such that the amplitude of the matter power spectrum as measured by the Cosmic 
Background Explorer (COBE) matches that obtained from the X-ray cluster 
abundance, the decrease in the predicted lens abundance is less than 25\% for 
$-1 \le w < -0.4$. We show that a more promising method for constraining the 
dark energy---one that is largely unaffected by the $\Omega_0 - w$ degeneracy 
as well as uncertainties in observational noise---is to compare the relative 
abundance of virialized X-ray lensing clusters with the abundance of 
non-virialized, X-ray underluminous, lensing halos. For aperture sizes of 
$\sim 15$ arcmin, the predicted ratio of the non-virialized to virialized 
lenses is greater than 40\% and varies by $\sim 20$\% between $w=-1$ and 
$w = -0.6$. Overall, we find that if all other weak lensing parameters are 
fixed, a survey must cover at least $\sim 40$ square degrees in order for the 
weak lens number count to differentiate a $\Lambda$CDM cosmology from a 
dark-energy model with $w=-0.9$ at the 3$\sigma$ level. If, on the other hand, 
we take into account uncertainties in the lensing parameters, then the 
non-virialized lens fraction provides the most robust constraint on $w$, 
requiring $\sim 50$ square degrees of sky coverage in order to differentiate 
a $\Lambda$CDM model from a $w=-0.6$ model to 3$\sigma$.
\end{abstract}


\section{INTRODUCTION}

Observations of distant type Ia supernovae (SNIa) indicate that the universe 
is undergoing a phase of accelerated expansion (Perlmutter et al. 1999, Riess 
1998). This, combined with the flat geometry favored by the cosmic microwave 
background (CMB) measurements (Miller et al. 1999, de Bernardis et al. 2002, 
Halverson et al. 2002, Sievers et al. 2002, Lee et al. 2001) and the evidence 
for a low matter-density with $\Omega_0 \sim 0.3$ (Peacock 2001, Percival et al. 
2001), suggests that the bulk of the total energy density of the universe is in 
the form of some exotic dark energy with a negative equation of state. One of the 
primary objectives of cosmology today is to uncover the origin and nature of 
this dark energy. 

A possible candidate for the dark energy is a cosmological constant $\Lambda$, 
with an equation of state $w = p / \rho$ (where $p$ is the pressure and $\rho$ 
is the energy density of the dark energy) strictly equal to $-1$. Another 
possibility, and one that may find favor from a particle-physics point of view, is 
a dynamical scalar field, termed quintessence, $Q$. Unlike the cosmological constant, 
the $Q$-component is both time-dependent and spatially inhomogeneous with an 
equation of state $w > -1$ that is likely to be redshift dependent. Determining 
the value of $w$ and how it changes with time are key to constraining 
the nature of the dark energy. 

While the accelerating expansion implies only that $w < -1/3$, combinations of CMB 
data, SNIa data, and large-scale-structure data suggest that $w$ is most likely in 
the range $-1 \le w < -0.6$ (Wang et al. 2000, Huterer \& Turner 2001, Bean \& 
Melchiorri 2002, Baccigalupi et al. 2002). Though combining these different data 
sets have provided some constraint on $w$, how $w$ should vary with redshift is 
largely unknown. Particle physics offer several possible functional forms 
for the quintessence field's potential $V(Q)$ and hence possible scenarios for 
the time history of $w$. Nonetheless, determining $w$'s redshift evolution 
observationally is likely to be very challenging (Barger \& Marfatia 2001,
Maor et al. 2001, Weller \& Albrecht 2001).

Strengthening the measured constraint on $w$ and perhaps excluding the 
cosmological constant as the source of the dark energy appear, however,
to be attainable goals within the near future. Since the dark-energy dynamics 
influences both the evolution of the background cosmology and the growth of 
structure, it directly affects many observables. Its modification of the 
angular-diameter 
distance, the luminosity distance, and the amplitude of the matter power spectrum, 
are the primary sources of dark-energy constraint in measurements of CMB 
anisotropies, SNIa, and local cluster abundances, respectively.

In this paper we consider another possible means of constraining $w$: measurement 
of weak gravitational-lens abundances. Weak lensing---the weak distortion of 
background-galaxy images due to the deep gravitational potential 
of an intervening overdensity--- provides a powerful technique for 
mapping the distribution of matter in the universe (see reviews by Bartelmann 
\& Schneider 2001, Mellier 1999). 
Here we study the impact of the dark energy on the predicted redshift distribution 
and sky density of weak lenses. Dark energy affects the abundance of weak lenses 
by not only modifying the distance-redshift relation and the matter power spectrum
but also by altering the rate of structure growth. In particular, the larger $w$ is 
the faster and earlier objects collapse. An interesting consequences of this is that if 
we separate weak lenses into the two observational classes---those that have 
collapsed and reached virial equilibrium and are therefore X-ray luminous and those 
that are non-virialized and hence X-ray underluminous (Weinberg \& Kamionkowski 2002; 
hereafter WK02)---the abundance of one class evolves slightly differently from the other. 
Therefore the relative fraction of these two types of lenses varies with $w$. This
observable is especially promising as compared to measurements of absolute abundances 
because it is less sensitive to uncertainties in both the cosmological parameters and 
the noise in the lensing map. 
 
This paper is organized as follows. In Section 2 we briefly summarize the weak-lensing 
signal-to-noise estimator and discuss how we determine the mass- and redshift-dependent 
minimum overdensity required to produce a detectable weak-lensing signal. Section 3 is 
devoted to the spherical-collapse model in quintessence cosmologies. We provide fitting 
formulas for the non-linear overdensity at virialization and the linear-theory density 
at collapse and describe our approach to normalizing the matter power spectrum. In 
Section 4 we show the resulting effect the dark energy has on the weak-lens abundances 
and in Section 5 we present our conclusions.

Finally, we note that a similar analysis has recently been performed by 
Bartelmann, Perrotta \& Baccigalupi (2002), although not for the case of 
non-virialized lenses. Although we agree with their general 
conclusion that the weak lens abundance is a potentially sensitive probe of the 
dark energy, our results differ from their results in important details. We discuss 
these differences in Section 4.2.

\section{Minimum Overdensity Needed to Produce Detectable Lensing Signal}

In order to compute the abundance of weak gravitational lenses for 
dark-energy cosmologies we must first determine the necessary conditions 
for a halo of a given density profile and redshift to produce a detectable 
weak-lensing signal. Of course the more overdense a halo is relative to the 
background density the more
it coherently distorts the nearby background galaxies and hence the stronger 
its lensing signal. The detectability of this signal is hampered, however, 
by noise in the weak-lensing map, primary of which is the intrinsic 
ellipticity distribution of the background galaxies.  The goal is therefore 
to determine the minimum overdensity a halo must have such that it produces
a sufficiently large signal relative to the noise so as to be detectable. 
A convenient method for computing this minimum overdensity is provided by 
Schneider's (1996) aperture-mass technique. 

Consider a lens at redshift $z_{\rm{d}}$ of surface mass density 
$\Sigma(\vartheta)$ within an angular radius $\vartheta$. For a source at 
redshift $z_{\rm{s}}$ the convergence $\kappa$ is given by,
\begin{equation}
\kappa(\vartheta)=
\frac{\Sigma(\vartheta)}{\Sigma_{\rm
{crit}}}, \; \; \; 
\Sigma_{\rm{crit}}=\frac{c^2}{4 \pi G}\frac{D_s}{D_d D_{ds}},
\end{equation}
where $D_{\rm{d}}$, $D_{\rm{s}}$ and $D_{\rm{ds}}$ are the angular-diameter 
distances between the lens and the observer, the source galaxy and the 
observer, and the lens and the source, respectively. Following Schneider 
(1996), define a spatially-filtered mass inside a circular aperture of 
angular radius $\theta$,
\begin{equation}
M_{\rm{ap}}(\theta) \equiv \int d^2\mbox{\boldmath$\vartheta$\unboldmath} \,
       \kappa(\mbox{\boldmath$\vartheta$\unboldmath}) \, 
       U(|\mbox{\boldmath$\vartheta$\unboldmath}|),
\end{equation}
where $U(\vartheta)$ is a continuous weight function that vanishes for 
$\vartheta > \theta$. If $U(\vartheta)$ is a compensated filter function,
\begin{equation}
\int_0^\theta d\theta \, \mbox{\boldmath$\vartheta$\unboldmath} \,
                  U(\mbox{\boldmath$\vartheta$\unboldmath}) = 0,
\end{equation}
then $M_{\rm{ap}}$ can be expressed in terms of the tangential component of
the observable shear, $\gamma_{\rm{t}}$,
\begin{equation}
M_{\rm{ap}}(\theta) = \int d^2\vartheta \,
       \gamma_{\rm{t}}(\vartheta) \, 
       Q(|\vartheta|),
\end{equation}
where the function $Q$ is related to $U$ by
\begin{equation}
Q(\vartheta) = \frac{2}{\vartheta^2} \int_0^\vartheta d\vartheta' \,
       \vartheta' \, U(\vartheta') \, - U(\vartheta).  
\end{equation}
In this paper we use the $l = 1$ radial filter function from the family 
given in Schneider et al. (1998): 
\begin{equation}
U(\vartheta) =\frac{9}{\pi \theta^2}(1-x^2)(\frac{1}{3}-x^2), \; \; \;
Q(\vartheta) =\frac{6}{\pi \theta^2} x^2(1-x^2),  
\end{equation}
where $x=\vartheta / \theta$. Taking the expectation value over galaxy positions
and taking into account the redshift distribution of source galaxies then
gives,
\begin{equation}
M_{\rm{ap}}(\theta) = \langle Z \rangle \int d^2\vartheta \,
      \langle \gamma_{\rm{t}} \rangle(\vartheta) \, 
       Q(|\vartheta|),
\end{equation}
where $ \langle \gamma_t \rangle (\vartheta)$ is the mean tangential shear
on a circle of angular radius $\vartheta$. The function $\langle Z \rangle$,
given by,
\begin{equation}
\langle Z \rangle = \int dz_s \; p_z(z_s) Z(z_s;z_d),
\end{equation}
where $p_z(z_s)$ is the redshift distribution of source galaxies and 
(Seitz \& Schneider 1997)
\begin{equation}
Z(z_s; z_d) \equiv
\frac{\lim_{z_s \rightarrow \infty}
\Sigma_{\rm{crit}}(z_d;z_s)}{\Sigma_{\rm{crit}}(z_d;z_s)}=
\frac{\Sigma_{\rm{crit}_{\infty}}(z_d)}
{\Sigma_{\rm{crit}}(z_d;z_s)},
\end{equation}
allows a source with a known redshift distribution to be collapsed onto a single
redshift $z_s$ satisfying $Z(z_s)= \langle Z \rangle$ (Seitz \& Schneider 1997; 
Bartelmann \& Schneider 2001). The source-redshift distribution is taken to be, 
\begin{equation}
p_z(z_s)=\frac{\beta z_s^2}{\Gamma
(3/\beta)z_0^3}\exp\left[-(z_s/z_0)^{\beta}\right],
\end{equation}
with $\beta=1.5$ and mean redshift $\langle z_s \rangle \approx 1.5 z_0 = 1.2$
(cf., Smail et al. 1995; Brainerd et al. 1996; Cohen et al. 2000).
Finally, assuming the ellipticities of different 
images are uncorrelated it can be shown (cf., Kruse \& Schneider 1999) that 
the dispersion $\sigma_{\rm{M}} (\theta)$ of $M_{\rm{ap}}$ is
\begin{equation}
\sigma_{\rm{M}}^2(\theta) = \frac{\pi \sigma_{\epsilon}^2}{n}
\int_0^{\theta}
d\vartheta \, \vartheta \, Q^2(\vartheta),
\end{equation}
where $n$ is the number density of galaxy images and $\sigma_{\epsilon}$ is
the dispersion in the galaxies' intrinsic ellipticity. In this paper we assume
$n = 30$ arcmin$^{-2}$ and $\sigma_{\epsilon} = 0.2$. The signal-to-noise ratio 
$\sl{S}$ within an aperture radius $\theta$ is then given by,
\begin{equation}
\sl{S}=\frac{M_{\rm{ap}}}{\sigma_{\rm{M}}}
=\frac{2 \langle Z \rangle \sqrt{\pi n}}{\sigma_{\epsilon}}
\frac{\int_0^{\theta}
d\vartheta \, \vartheta \langle \gamma_t \rangle (\vartheta) \, Q(\vartheta)}
{\sqrt{\int_0^{\theta}
d\vartheta \, \vartheta \, Q^2(\vartheta)}}.
\end{equation}

The tangential shear at $\vartheta$, $\langle \gamma_t \rangle (\vartheta)$, 
depends on the amplitude and shape of the lensing halo's density profile. 
Bartelmann (1995) showed that  $\langle \gamma_t \rangle(\vartheta) 
= \bar{\kappa}(\vartheta)-\langle \kappa \rangle(\vartheta)$, where 
$\langle \kappa \rangle(\vartheta)$ is the
dimensionless mean surface mass density on a circle of radius $\vartheta$
and $\bar{\kappa}(\vartheta)$ is the dimensionless mean surface mass
density within a circle of radius $\vartheta$. In this paper we describe the
mass density of lensing halos with the universal density profile introduced
by Navarro, Frenk \& White (1996; 1997; hereafter NFW). Thus, for an NFW halo 
at a given redshift with a given mass and mean overdensity relative to the 
background ($\Delta \equiv  \langle \rho_{\rm{pert}} \rangle / \rho_{\rm{b}}$), 
we can solve for 
the parameters of the profile (i.e., the scale radius and the scale density)
and obtain an estimate of  $\langle \gamma_t \rangle (\vartheta)$. Details of 
how we solve for the NFW-profile parameters are given in the Appendix of WK02.
With the density profile known we can determine, using equation (12), the 
expected value of $\sl{S}$. The minimum mean overdensity, $\Delta_{\rm{min}}$,
needed to produce a detectable lens is then given by that overdensity for 
which $\sl{S} > \sl{S}_{\rm{min}}$. In this paper we assume 
$\sl{S}_{\rm{min}} = 5$ and $\theta = 5'$, unless stated otherwise.

\section{Spherical Collapse in Dark Energy Cosmologies}

\begin{figure}
\vspace{-2.3cm}
\centerline{\hbox{\hspace{2.9cm}} \psfig{file=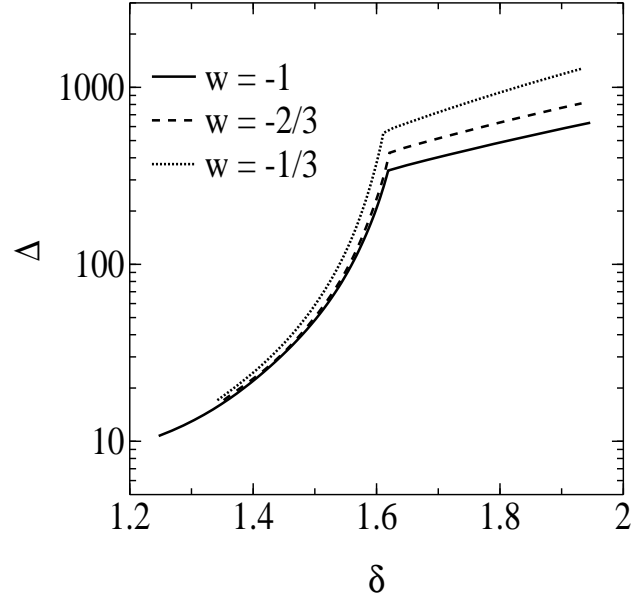,angle=0,width=3.4in}}
\vspace{1.2cm}
\caption{The nonlinear overdensity as a function of the linear-theory
overdensity for three constant-$w$ models. The full solution of the 
spherical-collapse model predicts collapse to an infinite overdensity as $\delta 
\rightarrow \delta_{\rm{c}} \sim 1.69$. According to the smoothing scheme, however, 
once a mass concentration reaches the virialization overdensity 
$\Delta_{\rm{vir}}(z)$, its radius remains constant so that the overdensity 
increases in proportion to the decrease in the background density. Shown are the 
smoothing-scheme solutions for mass concentrations that reach the virialization 
overdensity at $z=0$.}
\end{figure}
According to the spherical model of gravitational collapse a density 
perturbation with a nonlinear overdensity $\Delta$ corresponds to a particular 
position along the linear-theory evolutionary cycle.
Thus the minimum nonlinear overdensity $\Delta_{\rm{min}}$ described above 
corresponds to a minimum linear-theory overdensity $\delta_{\rm{min}}$; 
if an object of mass $M$ at redshift $z$ has a linear-theory overdensity 
$\delta > \delta_{\rm{min}} = \delta_{\rm{min}}(M,z)$, then it is sufficiently 
overdense to produce a detectable weak-lensing signal. By determining 
$\delta_{\rm{min}}$ from the computed $\Delta_{\rm{min}}$ we can apply the
Press-Schechter (1974) theory to calculate the number of halos per unit 
mass and redshift with $\delta > \delta_{\rm{min}}$ and hence 
$\sl{S} > \sl{S}_{\rm{min}}$. We can then find the redshift distribution and 
sky density of weak lenses and how these observables vary with $w$. We will show 
that for a broad range of dark-energy cosmologies a substantial fraction of 
detectable weak gravitational 
lenses have $\delta_{\rm{min}} < \delta_{\rm{c}} \approx 1.69$, where 
$\delta_{\rm{c}}$ is the critical density threshold for collapse. Those 
objects with $\delta < \delta_{\rm{c}}$ are commonly thought to be density 
perturbations that have not yet reached virialization and are therefore expected 
to have observational properties that are very different from typical 
virialized lensing clusters. 

In this Section, we present the approach used to map the minimum nonlinear 
overdensity $\Delta_{\rm{min}}$ to a minimum linear-theory overdensity 
$\delta_{\rm{min}}$ for quintessence models (QCDM). We describe the dynamical 
equations of gravitational collapse in QCDM and give fitting formulas for the 
nonlinear overdensity at virialization, $\Delta_{\rm{vir}}(z)$, and the 
critical density $\delta_{\rm{c}}$. We then discuss how we calculate the 
abundances of weak gravitational lenses, both those with 
$\delta < \delta_{\rm{c}}$ and those with $\delta > \delta_{\rm{c}}$.  
Below we assume a flat cosmology with a Hubble 
parameter $h = 0.65$, a spectral index $n_{\rm{s}} = 1$, a baryon density 
$\Omega_{\rm{b}}h^2 = 0.02$, and $\Omega_0 = 0.3$, unless stated otherwise.

\subsection{Dynamics}

\begin{figure}
\vspace{-2.3cm}
\centerline{\hbox{\hspace{2.9cm}} \psfig{file=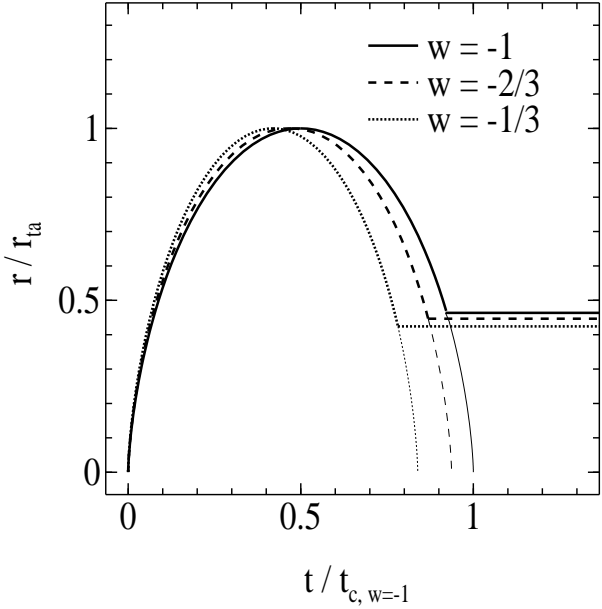,angle=0,width=3.4in}}
\vspace{1.2cm}
\caption{The radial evolution of a density perturbation that is collapsing today 
according to the spherical-collapse model. The ordinate gives the radius, $r$, 
in units of the turnaround radius, $r_{\rm ta}$, and the abscissa gives the time, 
$t$, in units of the overdensity collapse-time for the $\Lambda$CDM model.
As $w$ increases perturbations reach turnaround and collapse earlier, although growth is 
suppressed earlier as well. The collapse to a singularity 
predicted by the solution of the spherical-collapse model is avoided by 
the smoothing scheme (\emph{thick} curves) which yields a constant radius 
once the virialized overdensity is reached.}
\end{figure}
\begin{figure}
\vspace{-2.3cm}
\centerline{\hbox{\hspace{2.7cm}} \psfig{file=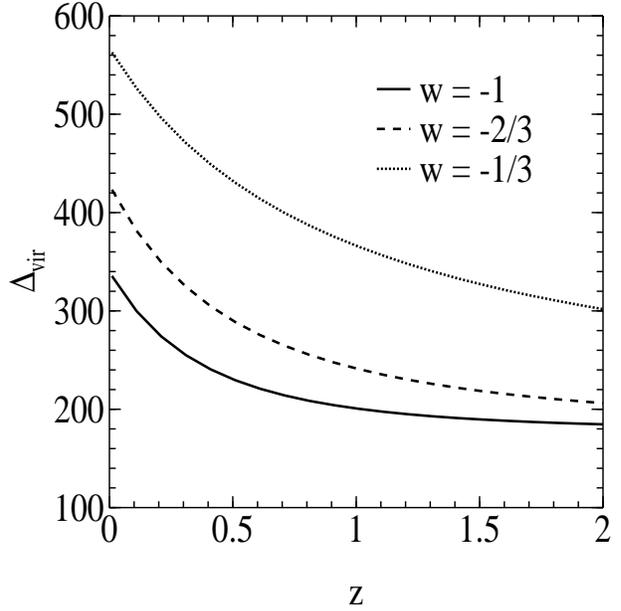,angle=0,width=3.4in}}
\vspace{1.2cm}
\caption{The non-linear overdensity at virialization as a function of redshift 
for three constant-$w$ models. As $w$ increases $\Delta_{\rm{vir}}$ increases 
because overdensities collapse earlier, when the mean gas temperature was higher.
For all models $\Delta_{\rm{vir}}$ asymptotes to the Einstein-de Sitter value of 
$178$ at high redshift.}
\end{figure}
In quintessence the dark energy is a dynamical, time-dependent component, Q, 
with an equation of state parametrized by $w \equiv p_{\rm{Q}} / \rho_{\rm{Q}}$, 
the pressure divided by the energy density. The evolution of the energy density 
with the cosmological scale factor goes as $\rho_{\rm{Q}} \propto a^{-3(1+w)}$, 
so that for $w = -1$ the standard cosmological-constant model, 
$\Lambda$CDM, is recovered. Current observational evidence cannot yet 
rule out a $w$ in the range $-1 \le w \la -0.5$. 

In order to relate a nonlinear overdensity to a linear-theory overdensity in 
QCDM we must first solve for the evolution of the overdensity's radius, $R$, 
with time. For a spherical overdensity patch with uniform matter density 
$\rho_{\rm{pert}} = 3 M / 4 \pi R^3$ the evolution is described by the momentum 
component of the Einstein equations (Wang \& Steinhardt 1998; hereafter WS98):
\begin{equation}
\frac{\ddot{R}}{R} = -\frac{4 \pi G}{3} 
              \left[ \rho_{\rm{pert}} + (1 + 3w) \rho_{\rm{Q}} \right].
\end{equation}
As WS98 pointed out, for $w \not= -1$ the space curvature $k_{\rm{pert}}$ 
inside the overdensity patch is time dependent. Physically, this is because the
evolution of the energy density in the $Q$ component is evolving independently
of the change in radius of the overdensity patch. As a result, one cannot assume 
that within the collapsing overdensity the rate of change of the internal energy 
in the Q-component, $u_{\rm{Q}}$, equals the rate of work done by the Q-component. 
That is, because $d\rho_{\rm{Q}}/dt$ is nonzero unless the Q-component is the 
cosmological constant,
\begin{eqnarray}
\frac{du_{\rm{Q}}}{dt} & = & \frac{d}{dt}\left(\rho_{\rm{Q}} V \right) \nonumber \\
          & \not= & - p_{\rm{Q}} \frac{dV}{dt},
\end{eqnarray}
where $V \propto R^3$ is the volume of the overdensity patch. Therefore equation 
(13) cannot be cast in the form of a first order differential equation as is 
often done when going from an acceleration equation to a Friedman-like energy
equation. Assuming a constant $k_{\rm{pert}}$, as was done in the version 1 
preprint of {\L}okas \& Hoffman (2001), yields significantly different solutions 
for the evolution of the radius, $R(t)$, and hence for $\Delta_{\rm{vir}}(z)$ 
and $\delta_{\rm{c}}$.

If we combine equation (13) with the Friedman equation for the background,
\begin{equation}
\left(\frac{\dot{a}}{a}\right)^2 = \frac{8 \pi G}{3} 
              \left( \rho_{\rm{b}}  +  \rho_{\rm{Q}} \right),
\end{equation}
and impose the boundary conditions $\left. dR/da \right|_{a = a_{\rm{ta}}} = 0$ and
$\left. R \right|_{a = 0} = 0$, where $a_{\rm{ta}}$ is the scale factor at turn-around, 
then for a spherical density perturbation with a 
given $\Delta$ and redshift $z$, the unique temporal evolution of the 
overdensity, from linearity to nonlinearity, can be solved (cf., Appendix A in 
WS98). We then have a one-to-one map from $\Delta(z)$ to $\delta(z)$, as shown 
in Figure 1 for the cases $w = -1, -2/3$ and $-1/3$. The map has a mild $w$ 
dependence, with a given $\delta$ corresponding to a slightly larger $\Delta$ 
as $w$ increases.
This is a consequence of the earlier formation of structure in QCDM models 
relative to $\Lambda$CDM models; overdensities collapse faster and are therefore 
more concentrated for $w > -1$. This point is well illustrated in Figure 2 where 
we show the growth of a spherical perturbation for the same quintessence models. 
As expected, the larger $w$ is, the earlier structures reach turnaround and collapse. 

It can be shown that in the limit $\delta \rightarrow \delta_{\rm{c}}$ the 
spherical-collapse model predicts that the radius, $R$, of the overdensity 
goes to zero and hence $\Delta \rightarrow \infty$. Of course well before 
reaching the singular solution an actual overdensity will virialize, thereby 
halting its collapse. To account for this fact we invoke a simple smoothing 
scheme in which the radius of the matter perturbation is constant with time 
upon reaching the virialized overdensity (see Figure 2). We refer the reader 
to WK02 for details of the smoothing method.

As described in WS98, the value of $\Delta_{\rm{vir}}(z)$ for quintessence models,
needed here in order to implement the smoothing scheme, can be obtained via the
virial theorem, energy conservation, and solving equations (13) and (15) for 
the overdensity at turnaround. In Figure 3 we show the resulting numerical solution 
to $\Delta_{\rm{vir}}(z)$. We find that an accurate fitting function to 
$\Delta_{\rm{vir}}(z)$ for $-1 \le w \le -0.3$, modeled after the approximation 
given in Kitayama \& Suto (1996) for a $\Lambda$CDM cosmology, is
\begin{equation}
\Delta_{\rm{vir}}(z) = 18 \pi^2 \left[ 1 + a \Theta^b(z) \right],
\end{equation}
where 
\begin{eqnarray}
a & = & 0.399 - 1.309 (|w|^{0.426} - 1), \nonumber \\
b & = & 0.941 - 0.205 (|w|^{0.938} - 1),
\end{eqnarray}
and $\Theta(z) = 1/\Omega_{\rm{m}}(z)-1 = (1/\Omega_0-1)(1+z)^{3w}$. Since structures 
start to form earlier the larger $w$ 
is, the mean gas temperature in collapsing objects is higher in larger-$w$ models. As
a result, a greater overdensity is required in order for such objects to become 
bound and virialized, explaining why $\Delta_{\rm{vir}}$ rises with increasing $w$. 
Note, however, that for $\Delta(z) < \Delta_{\rm{vir}}(z)$ the map 
from nonlinear to linear overdensity has a weak dependence on not only $w$ but on 
$\Omega_0$ and redshift as well. The critical threshold for collapse today 
$\delta_{\rm{c}}(z=0) = \delta_{\rm{c}}(z) \, D(0,\Omega_0,w)/D(z,\Omega_0,w)$,
where $D(z,\Omega_0,w)$ is the linear growth factor (see WS98), also has a weak dependence 
on $\Omega_0$ and $w$, as shown in Figure 4. For $0.1 \le \Omega_0 \le 1$ and 
$-1 \le w \le -0.3$, we find that an accurate fitting function to $\delta_{\rm{c}}(z)$,
also modeled after the approximation given in Kitayama \& Suto (1996) for a 
$\Lambda$CDM cosmology, is
\begin{eqnarray}
\delta_{\rm{c}}(z) & = & \frac{3(12 \pi)^{2/3}}{20} 
                       \left[ 1+\alpha \, \log_{10} \Omega_{\rm{m}}(z) \right], \nonumber \\
\alpha & = & 0.353 w^4  +  1.044 w^3  +  1.128 w^2 \nonumber \\
        &   &  + \,  0.555 w  +  0.131.
\end{eqnarray}
Incorrectly assuming that $k_{\rm{pert}}$ 
is constant, however, yields a $\delta_{\rm{c}}(z=0)$ with a much stronger dependence  
on these parameters, with inferred values for $\Omega_0 = 0.3$ of 
$\delta_{\rm{c}}(z = 0) \sim 1.5$ and $ \sim 1.0$ for $w = -2/3$ and $w = -1/3$, 
respectively ({\L}okas \& Hoffman 2001).    
\begin{figure}
\vspace{-2.3cm}
\centerline{\hbox{\hspace{2.9cm}} \psfig{file=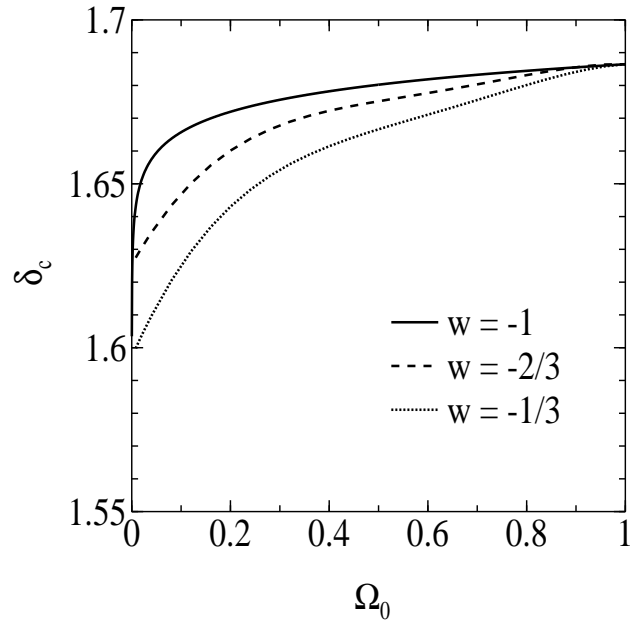,angle=0,width=3.4in}}
\vspace{1.2cm}
\caption{The linear-theory critical threshold for collapse, $\delta_{\rm{c}}$, at 
$z=0$ as a function of $\Omega_0$ for three constant-$w$ models. $\delta_{\rm{c}}$ 
does not vary significantly over a wide range in $w$ or $\Omega_0$. }
\end{figure}

\subsection{Abundances}

Since we are interested in computing the abundances of both virialized 
weak lenses and non-virialized weak lenses we consider two ranges of 
overdensity in our lens-abundance calculations:
(1) $\delta_{\rm{min}}< \delta < \delta_{\rm{c}}$, the non-virialized lenses,
and,
(2) $\delta > \delta_{\rm{c}} \ge \delta_{\rm{min}}$, the virialized lenses.  
As we showed in WK02, both lens types are cluster-mass overdensities. However,
while the virialized lenses are typically virialized clusters that form at 
rare (e.g., $ > 3\sigma$) high-density peaks of a Gaussian primordial distribution,  
the non-virialized lenses correspond to proto-clusters (e.g., $2\sigma - 3\sigma$ 
peaks)---mass overdensities that have not yet undergone gravitational collapse and 
virialized, but which have begun to break away from the cosmological expansion. 
These proto-clusters should contain galaxies and perhaps a few groups that later 
merge to form the cluster (cf., White et al. 2002). The timescale for collapse of 
cluster-mass objects is large, and the overdensities can be 
very large even before they have virialized.  It should therefore not be too surprising 
that proto-clusters produce a weak-lensing signal that resembles that from virialized 
clusters. 

Though the lensing signals may be similar, the two lens types are 
expected to have different observational features. In particular, since the X-ray 
luminosity is a very rapidly varying function of the virialized mass, the summed X-ray 
emission from a non-virialized lens should be much smaller than that from a fully 
virialized lensing cluster of the same mass. In referring to these proto-clusters as 
``dark'', we thus mean that they should be X-ray underluminous. Although the mass-to-light 
ratio of these clusters should be comparable to those for ordinary clusters, since 
(1) high-redshift clusters may be difficult to pick out in galaxy surveys and, (2) 
proto-clusters will typically have a sky density a few times smaller than ordinary clusters, 
it would also not be 
surprising if these dark lenses had no readily apparent corresponding galaxy overdensity. 
Observational evidence of such dark lenses has been reported in detections by Erben et al. 
(2000), Umetsu \& Futamase (2000), Miralles et al. (2002), Dahle et al. (2002), and 
Koopmans et al. (2000), the latter 
involving a detection through strong, rather than weak, lensing. A more detailed discussion 
of the features that may distinguish dark and virialized weak lenses is given in WK02.

In order to compute the abundances of virialized and dark lenses we need to know the 
probability that an object of a given mass at a given redshift is in one of the above 
mentioned ranges in overdensity. If we assume Gaussian statistics for the initial 
linear-theory density field, then the probability that an object's overdensity is in 
the range $\delta_1 < \delta < \delta_2$ is
\begin{equation}
P(\delta_1 < \delta < \delta_2) =
\mbox{erf}\left(\frac{\nu_2}{\sqrt 2
}\right) -
\mbox{erf}\left(\frac{\nu_1}{\sqrt 2}\right),
\end{equation}
where `erf' is the error function, $\nu = \delta / \sigma$, and  
$\sigma=\sigma (M,z)$ is the rms density fluctuation of an object of mass $M$ at 
redshift $z$. From Press-Schechter theory, we know that  
the comoving number density of virialized objects (those with $\delta > 
\delta_{\rm{c}}$) of mass $M = 4 \pi R^3 \rho_0 /3 $ in the interval $dM$ that are 
at redshift $z$ in a Universe with comoving background density $\rho_0$ is,
\begin{eqnarray}
\frac{dn}{dM}(M,z)=\sqrt{\frac{2}{\pi}} \frac{\rho_0}{M^2}
\frac{\delta_{\rm{c}}(z)}{\sigma(M,z)} \left|\frac{d \ln\sigma(M,z)}{d \ln M}\right| 
   \nonumber \\* \times  \exp \left[-\frac{\delta_{\rm{c}}(z)^2}{2 \sigma^2(M,z)} \right].
\end{eqnarray}
We can therefore compute the abundance of objects in the overdensity range 
$\delta_1 < \delta < \delta_2$ by convolving the above mass function of virialized objects 
with $P(\delta_1 < \delta < \delta_2) / P( \delta > \delta_{\rm{c}})$. Specifically, 
the fraction of objects that can lens relative to those that are virialized is, for 
dark lenses,
\begin{eqnarray}
f_{\rm{dark}}(M,z) & = & \left\{
\begin{array}{ll} \frac{{\textstyle P(\delta_{\rm{min}} < \delta <
\delta_{\rm{c}})}} {{\textstyle P(\delta >\delta_{\rm{c}})}},       & 
\delta_{\rm{min}} < \delta_{\rm{c}};\\ \\
 0, &   \mbox{otherwise},\end{array} \right.  
\end{eqnarray}
and for virialized lenses,
\begin{eqnarray}
f_{\rm{vir}}(M,z) & = & \left\{
\begin{array}{ll} \frac{\textstyle{ P(\delta > \delta_{\rm{min}})}}
 {{\textstyle P(\delta>\delta_{\rm{c}})}},     & \; \; \; \; \;\; \;
\delta_{\rm{min}} > \delta_{\rm{c}};\\ \\ 1, 
& \; \; \; \; \;\; \; \mbox{otherwise}.\end{array} \right.
\end{eqnarray}
As noted in WK02, the lower the mass of the object the larger the minimum 
overdensity needed to produce a detectable weak-lensing signal. For low 
enough masses the minimum overdensity becomes so large that both 
$f_{\rm{dark}}$ and $f_{\rm{vir}}$
approach zero, thereby imposing an effective weak-lensing mass threshold.
Given $f$ and equation (20) we can compute the total comoving number density 
of weak lenses of a particular type. Multiplying by the comoving volume 
element $dV_{\rm c} / \, dz \, d\Omega (w)$ then gives the differential number 
count of lensing objects per steradian, per unit redshift interval:  
\begin{equation}
\frac{dN}{dz \, d\Omega} = \frac{dV_{\rm c}}{dz \, d\Omega} 
\int_0^{\infty} f(M) \frac{dn}{dM}(M)dM.
\end{equation}
By integrating over redshift we can then compute the number of dark and virialized 
lenses we expect to see per unit area of sky for a given QCDM model. 

\subsection{Normalizing the power spectrum}

\begin{figure}
\vspace{-2.3cm}
\centerline{\hbox{\hspace{2.5cm}} \psfig{file=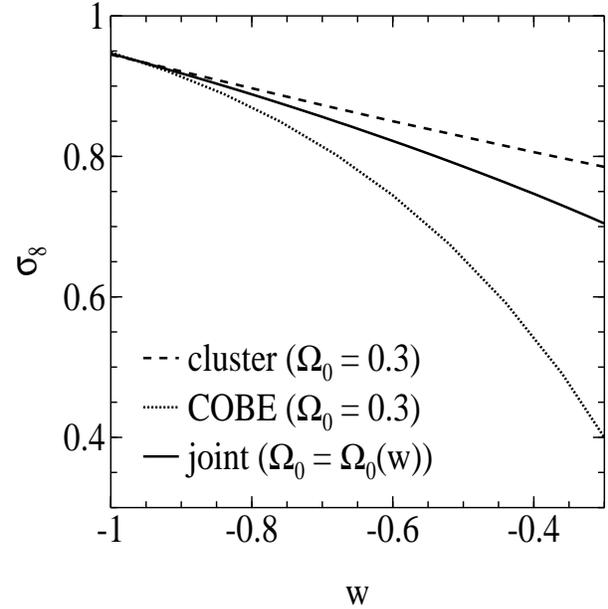,angle=0,width=3.4in}}
\vspace{1.2cm}
\caption{The dependence of $\sigma_8$ on $w$ as obtained using three different 
approaches: fixing $\Omega_0 = 0.3$ and normalizing to the observed X-ray cluster 
abundance (\emph{dashed} line), fixing $\Omega_0 = 0.3$ and normalizing to COBE 
(\emph{dotted} line), and allowing $\Omega_0$ to vary with $w$ such that the cluster 
abundance constraint matches the COBE constraint (\emph{solid} line).}
\end{figure}
\begin{figure}
\vspace{-1.3cm}
\centerline{\hbox{\hspace{2.5cm}} \psfig{file=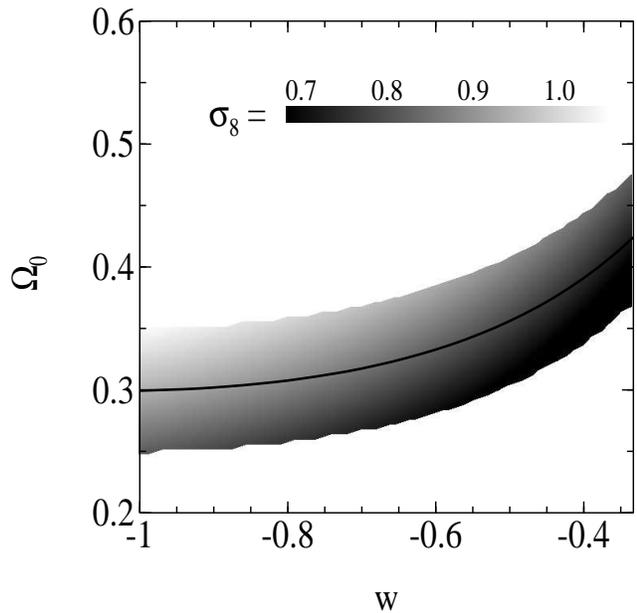,angle=0,width=3.4in}}
\vspace{1.2cm}
\caption{The region in the $\Omega_0-w$ plane where the X-ray cluster abundance 
constraint of $\sigma_8$, at 95\% confidence, overlaps the COBE constraint of $\sigma_8$. 
The gray scale gives the corresponding $\sigma_8$ values and the \emph{solid} line shows 
wheres the central values match.}
\end{figure}
\begin{figure*}
\vspace{-1.7cm}
\centerline{\hbox{\hspace{2.0cm}} \psfig{file=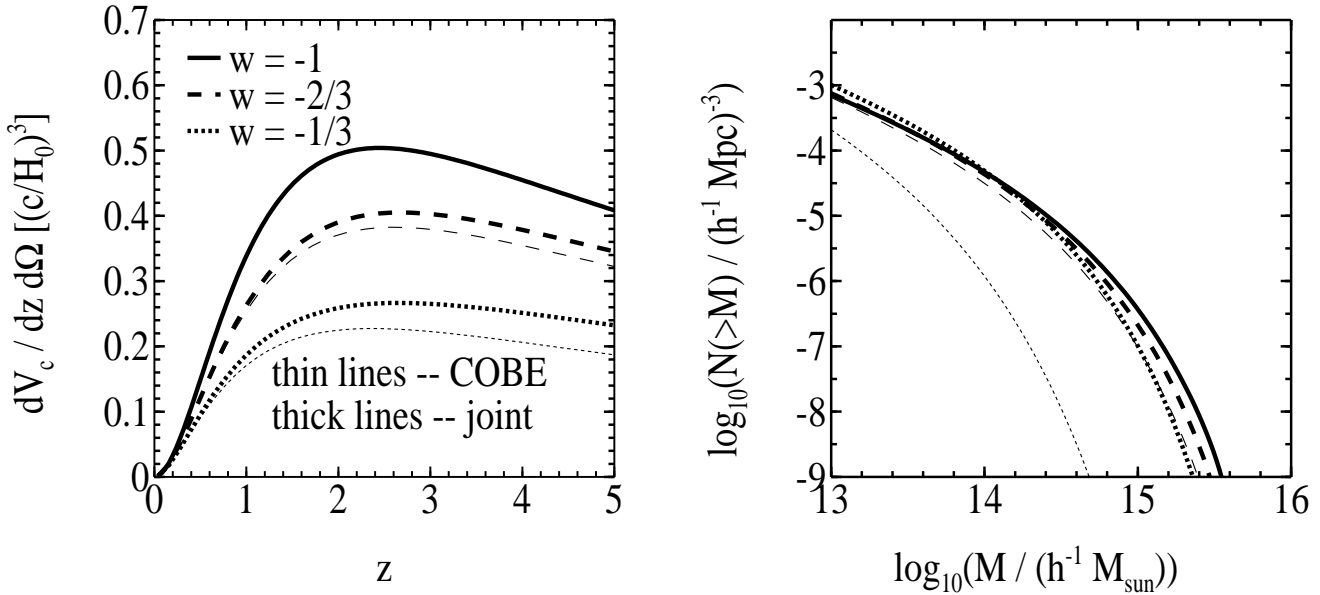,angle=0,width=6.6in}}
\vspace{1.2cm}
\caption{The comoving volume element as a function of redshift (\emph{left} panel) and 
the comoving number density of virialized objects as a function of mass 
(\emph{right} panel)
for three constant-$w$ models. Results are shown for both the COBE normalization 
of $\sigma_8$ with fixed $\Omega_0=0.3$ (\emph{thin} lines) and the joint 
cluster abundance--COBE normalization of $\sigma_8$ with $\Omega_0 = \Omega_0(w)$ 
(\emph{thick} lines).}
\end{figure*}
In equation (23) the volume term and the two terms within the integrand are all 
functions of $w$. While the predicted abundance of weak lenses will therefore vary 
with $w$, the degree to which it will vary depends on the shape and normalization
of the power spectrum of density fluctuations. In particular, to compute the abundance 
of weak-lenses we need to know $\sigma(M,z)$. 

For the {\it shape} of the power spectrum we use the fitting formulas given in 
Ma et al. (1999) for QCDM models with the transfer function and shape parameter 
for $\Lambda$CDM models given by Bardeen et al. (1986) and Hu \& Sugiyama (1996, 
eqs. [D-28] and [E-12]), respectively. Since the Q-component does not cluster on 
scales less than $\sim 100$ Mpc (Caldwell, Dave \& Steinhardt 1998), at the 
weak-lensing scales the shape of the spectrum does not differ significantly from 
the well studied $\Lambda$CDM shape. 

The {\it normalization} of the power spectrum, often expressed in terms of $\sigma_8$,
the rms fluctuation today at a scale of 8 $h^{-1}$ Mpc, is not as well
constrained as its shape and will in general be a function of $w$. There are two 
different methods commonly used to obtain the normalization: to fix it by the observed 
X-ray cluster abundance or to fix it by the CMB 
large-scale anisotropies observed by the COBE satellite. Both approaches have comparable 
uncertainties; the cluster abundance constraint on $\sigma_8$ has a 20\% uncertainty
at the 2$\sigma$ level (WS98) while the COBE constraint has a 7\% uncertainty at 
the 1$\sigma$ level (Bunn \& White 1997). To obtain an estimate of 
how $\sigma_8$ varies with $w$ so that we may, in turn, determine how 
$dN / dz d\Omega$ varies with $w$ for dark and virialized lenses, we will 
consider three possible approaches. The first two involve fixing the cosmological 
parameters (e.g., $\Omega_0$, $h$, $\Omega_{\rm{b}}$, $n_{\rm{s}}$) and using 
either the cluster-abundance constrained $\sigma_8(w)$ or the COBE constrained 
$\sigma_8(w)$. For the former we will use the fit given in WS98 and for the 
latter the fit given by Ma et al. (1999); see Figure 5. The third approach is to 
allow the cosmological parameters to be free parameters and then jointly match 
the cluster-abundance constraint with the COBE constraint so that each gives the same 
$\sigma_8(w)$.  Since measurements of $\sigma_8$ are most degenerate with $\Omega_0$,
we will let $\Omega_0$ be the parameter that varies. In Figure 6 we show the 
region in the $\Omega_0$--$w$ plane where the X-ray cluster-abundance constraint,
at the 95\% confidence level, overlaps the COBE constraint. The solid curve shows 
where the central values match, with the resulting range in $\Omega_0$  
($0.3 \la \Omega_0 \la 0.4$ for $-1 < w < -0.4$) within observational uncertainties 
(Wang et al. 2000). The corresponding $\sigma_8(w)$ curve is shown in Figure 5. As 
we will show, the predicted weak-lens abundances and how they vary with $w$ strongly 
depend on which $\sigma_8(w)$ normalization approach is chosen.

\section{Results}

\begin{figure*}
\vspace{-1.5cm}
\centerline{\hbox{\hspace{2.0cm}} \psfig{file=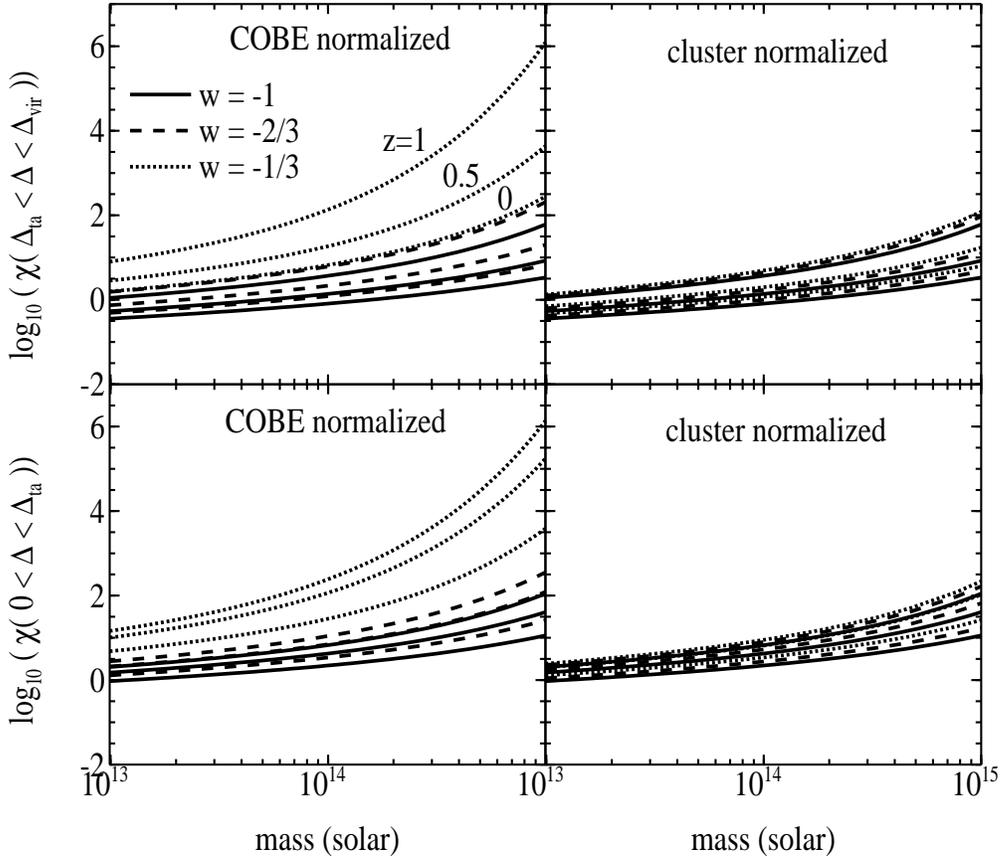,angle=0,width=5.3in}}
\vspace{1.2cm}
\caption{The fraction $\chi$ of objects with overdensities in the range 
$\Delta_{\rm{ta}} < \Delta < \Delta_{\rm{vir}}$ (\emph{upper} panels) and 
$0 < \Delta < \Delta_{\rm{ta}}$ (\emph{lower} panels) relative to those objects 
that are virialized, $\Delta > \Delta_{\rm{vir}}$, as a function of mass. 
The \emph{left} panels correspond to the COBE normalization of $\sigma_8$ with
$\Omega_0=0.3$ and the \emph{right} panels correspond to the X-ray cluster abundance 
normalization of $\sigma_8$ with $\Omega_0=0.3$. For each constant-$w$ model we show  
the fraction $\chi$ at $z=0$ (bottom curve), $z=1/2$ (middle curve), and $z=1$ 
(top curve).}
\end{figure*}
We are interested in determining whether the number count and redshift distribution 
of both dark and virialized weak lenses have the potential to constrain $w$. Another possibly 
useful observable for this purpose is the number count of dark lenses relative to virialized 
lenses. Since dark lenses are at an earlier stage of their dynamical evolution as compared 
to virialized lenses, those cosmologies that favor a faster growth of structure 
(i.e., QCDM models with larger $w$) will, for a given $\sigma_8$, have fewer dark lenses and 
more virialized lenses. The ratio of the two is therefore expected to vary with $w$. 
A priori, this latter observable seems particularly promising. As discussed in WK02, the ratio 
of dark to virialized lenses is not very sensitive to observational noise in the weak lensing 
maps since observational noise equally affects the detectability of both types of lenses. 
Contrastingly, uncertainties in observational noise will make it difficult to constrain $w$ by 
simply comparing predicted weak-lens number counts with observed weak-lens number counts. 

Before presenting how the above observables are modified by the dark energy we first discuss 
how each of the factors that determine the observed abundance are affected by changes in $w$. 
Doing so provides both physical insight into the results and illustrates the calculational 
procedure discussed in the previous Sections. 

\subsection{Preliminaries}

As noted above, the predicted abundance of weak lenses will vary with $w$ on account of three 
factors: the comoving volume element, the Press-Schechter comoving number density of virialized 
objects, and the value of $f_{\rm{dark/vir}}$ [equations (21) and (22)]. The degree to which 
each varies depends on the chosen $\sigma_8(w)$  normalization.  As Figure 7 shows, 
$dV_{\rm{c}} / dz \, d\Omega$ decreases monotonically with increasing $w$ for both fixed 
$\Omega_0$ and $\Omega_0=\Omega_0(w)$ as given by jointly normalizing $\sigma_8$ to COBE 
and the cluster abundance. However, because the joint normalization yields a larger $\Omega_0$ 
with $w$ and a less significant decline in $\sigma_8$ for $w > -1$ as compared to the 
COBE normalization with $\Omega_0$ fixed, the former approach predicts a nearly constant 
virialized object number density with increasing $w$ while the latter predicts a significant 
decrease in the number density. 
\begin{figure*}
\vspace{-2.0cm}
\centerline{\hbox{\hspace{2.0cm}} \psfig{file=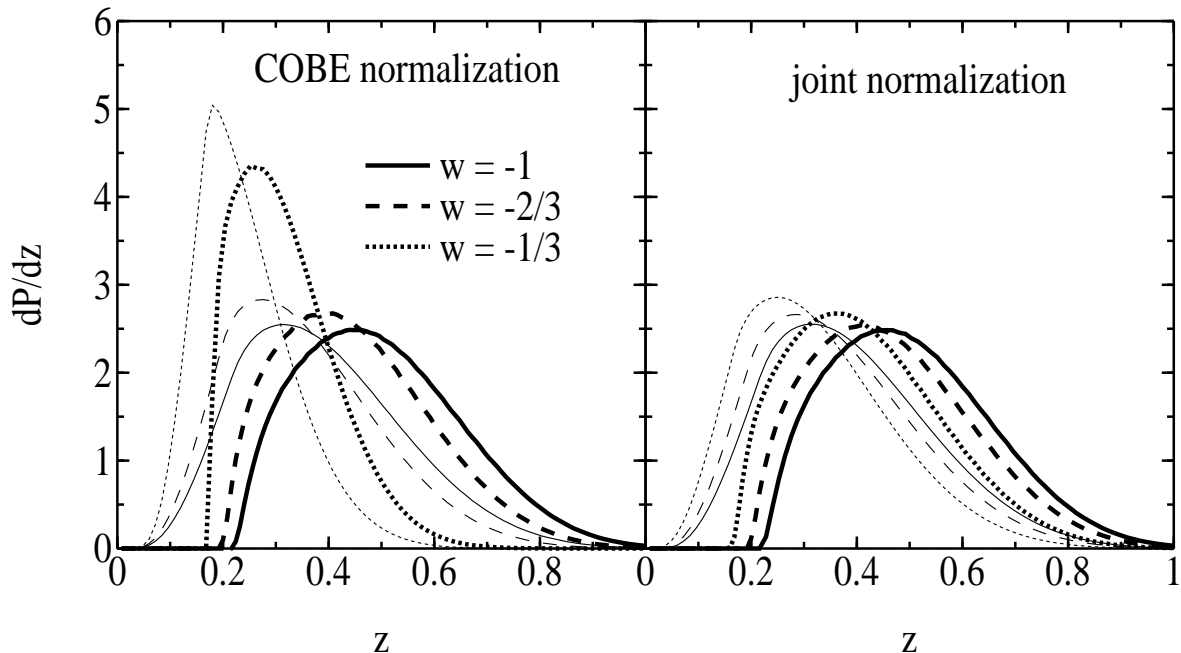,angle=0,width=6.4in}}
\vspace{1.2cm}
\caption{The normalized redshift distribution of virialized lenses (\emph{thin} lines) 
and non-virialized lenses (\emph{thick} lines) for three constant-$w$ models. The 
\emph{left} panel shows results obtained when $\sigma_8$ is normalized to COBE with 
$\Omega_0 = 0.3$ and the \emph{right} panel when $\sigma_8$ 
is jointly normalized to COBE and the X-ray cluster abundance with $\Omega_0 = \Omega_0(w)$.
The peak of the redshift distributions shift toward lower redshifts as $w$ increases 
because $\sigma_8$ decreases with $w$. The shift in the peaks is less drastic, however, when the 
joint normalization is assumed.}
\end{figure*}

A similar trend is seen in the functions $f_{\rm{vir}}$ and 
$f_{\rm{dark}}$, as Figure 8 demonstrates. Here we plot the fraction of objects that have not yet 
reached turnaround ($0 < \Delta < \Delta_{\rm{ta}}$) and the fraction of objects that are between 
turnaround and virialization ($\Delta_{\rm{ta}} < \Delta < \Delta_{\rm{vir}}$) relative to those
objects that are virialized ($\Delta > \Delta_{\rm{vir}}$). The figure illustrates several key 
elements of structure formation according to the spherical-collapse model for dark-energy 
cosmologies. First, the fraction, $\chi$, of objects in both of these lower-overdensity 
ranges increases with mass in accordance with the hierarchical growth of structure. The 
fraction also increases with 
redshift since objects are collapsing and evolving toward virialization. It is also interesting 
to note that objects with $\Delta_{\rm{ta}} < \Delta < \Delta_{\rm{vir}}$ evolve more rapidly as 
compared to objects with $0 < \Delta < \Delta_{\rm{ta}}$. This is demonstrated by the fact that 
at $z=1$ the fraction of both types of objects is nearly the same though by $z=0$ there are more 
objects that have not reached turnaround. Furthermore, the larger $w$ is, the greater the 
difference between the rates of evolution. These effects are a consequence of the 
suppression of structure growth in cosmologies with dark energy; namely, growth slows down 
earlier for larger $w$ and those objects that are less overdense at a given redshift have 
greater difficulty overcoming the repulsive effects of the dark energy and collapsing. Finally, 
the plots show how strongly the fraction depends on the chosen $\sigma_8$ normalization, with a 
significant variation with $w$ for the COBE normalization and a fairly small variation for the 
cluster-abundance normalization. This, in turn, means that the degree to which the functions 
$f_{\rm{vir}}$ and $f_{\rm{dark}}$ vary with $w$ is highly dependent on the assumed normalization 
approach.

\subsection{Weak lens abundances}
In Figure 9 we show the predicted redshift distribution of virialized lenses and dark lenses 
for three constant $w$ models. For the COBE normalized $\sigma_8$ with fixed $\Omega_0$ the 
distributions show a fairly strong sensitivity to $w$. As $w$ increases from $-1$ to $-1/3$ 
the peak of the distributions shift toward lower redshifts. Although one might expect the 
trend to be in the opposite direction given that structures form faster for larger $w$ models, 
the effect is counteracted by the decrease in $\sigma_8$ with increasing $w$. That the decrease 
in $\sigma_8$ so overwhelms any tendency for structure to form faster for $w>-1$ is not 
surprising given the weak $w$ dependence in the $\Delta$ -- $\delta$ map (Figure 1) and in the
function $\delta_{\rm{c}}(z)$ (Figure 4). Note, however, that the shift in the distributions 
with $w$ becomes much less significant if a joint COBE -- cluster abundance normalization is 
assumed. Finally, given that dark lenses are likely progenitors of virialized clusters, it 
is not surprising 
that both normalization approaches predict that the dark lenses have a larger mean redshift 
than the virialized lenses. 

To determine how well the weak lens redshift distributions can constrain $w$ we generated mock 
redshift data and determined (using the Kolmogorov-Smirnov test) the probability of 
differentiating two 
different constant-$w$ models as a function of the number of lenses detected. We found that 
to differentiate a $\Lambda$CDM model from both a $w=-0.6$ model and a $w=-0.9$ model at the 
3$\sigma$ level required, on average, approximately 200 weak lenses and 2000 weak lenses, 
respectively. As we show below, this corresponds to a survey coverage of $\sim 15$ and $\sim 150$ 
square degrees. Note, however, that for sufficiently wide surveys systematic uncertainties such 
as mass-redshift selection effects and lens density profiles might dominate the errors. 

By integrating over the redshift distribution we obtain the total number of virialized and 
dark lenses expected per square degree on the sky. As Figure 10 shows, the COBE normalization 
with $\Omega_0=0.3$ shows a significant decline in the number count as $w$ increases. By 
$w=-2/3$ the number count of both virialized and dark lenses has dropped by a factor of two 
from the $\Lambda$CDM value. The joint normalization, in which we allow $\Omega_0$ to vary 
with $w$, predicts a much more mild dependence on $w$ with the number count dropping by 
only $\sim 20$\%  from $w=-1$ to $w=-2/3$ for both lens types. Therefore, while the COBE-only 
normalization approach predicts that the sky coverage 
needed to distinguish the $\Lambda$CDM model from a $w=-0.6$ model to 3$\sigma$ is only $\sim 2$ 
degree$^2$, the joint approach requires $\sim$ 15 degree$^2$. Similarly, to distinguish the 
$\Lambda$CDM model from $w=-0.9$ requires $\sim 40$ degree$^2$ and $\sim 100$ degree$^2$, 
respectively. The systematic uncertainties affecting absolute sky density measurements, such 
as noise in the lensing maps and uncertainties in the lens density profiles, are expected to 
add further complications. This suggests that it will be very difficult to constrain $w$ 
using just the number count of either virialized or dark lenses without, at the very 
least, a tighter constraint on $\Omega_0$.
\begin{figure*}
\vspace{-3.2cm}
\centerline{\hbox{\hspace{4.0cm}} \psfig{file=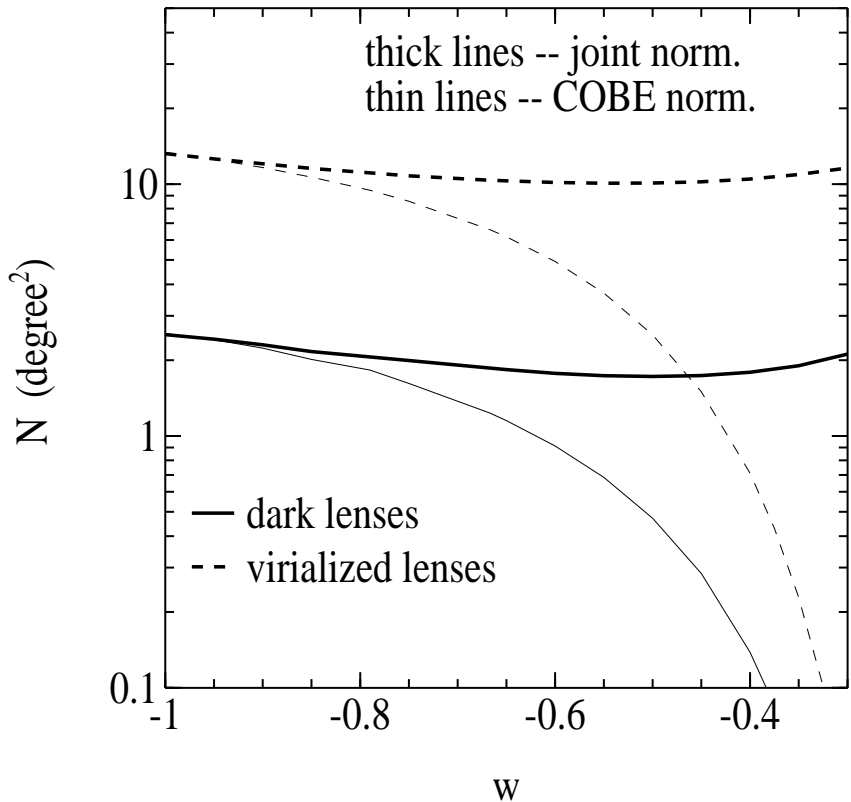,angle=0,width=4.7in}}
\vspace{1.2cm}
\caption{The total number of virialized lenses (\emph{dashed} curves) and non-virialized 
lenses (\emph{solid} curves) per square degree as a function of $w$. \emph{Thin} lines 
correspond to the COBE normalized $\sigma_8$ with $\Omega_0=0.3$ and \emph{thick} lines 
to the joint COBE--cluster abundance normalized $\sigma_8$ with $\Omega_0=\Omega_0(w)$. While 
the number count drops by a factor of two between $w=-1$ and $w=-2/3$ for the COBE-only
normalization, the drop is much less significant for the joint normalization.}
\end{figure*}
 
We also note that our results do not agree with the results found by 
Bartelmann, Perrotta \& Baccigalupi (2002; hereafter BPB). They 
found that from $w=-1$ to $w \approx -0.6$, the number of virialized weak lenses per square 
degree {\it increases} by nearly a factor of two. The increase is roughly linear up to 
the maximum after which the number count declines steeply. In obtaining these results, 
however, they use the formulas for $\Delta_{\rm{vir}}$ 
and $\delta_{\rm{c}}$ given in {\L}okas \& Hoffman (2001) who assume that the space curvature 
within a collapsing overdensity patch is time-independent. As we showed in Section 3.1, this
assumption is invalid for $w \not= -1$ and leads to incorrect values for $\Delta_{\rm{vir}}$ 
and $\delta_{\rm{c}}$. To confirm that this is the source of our differences, we recomputed 
the number count of weak lenses as a function of $w$ using the algorithm described in BPB
(which differs somewhat from ours because we are interested in separating lenses into
virialized and non-virialized types). When we assume the incorrect 
{\L}okas \& Hoffman (2001) values 
for $\Delta_{\rm{vir}}$ and $\delta_{\rm{c}}$ we recover the results found by BPB; 
however, when we assume the values for  $\Delta_{\rm{vir}}$ and $\delta_{\rm{c}}$ predicted by 
solving the spherical-collapse equations of Section 3.1, we obtain results very similar to those 
described in the preceding paragraphs. 

\subsection{Fraction of lenses that are dark}

\begin{figure}
\vspace{-2.3cm}
\centerline{\hbox{\hspace{2.7cm}} \psfig{file=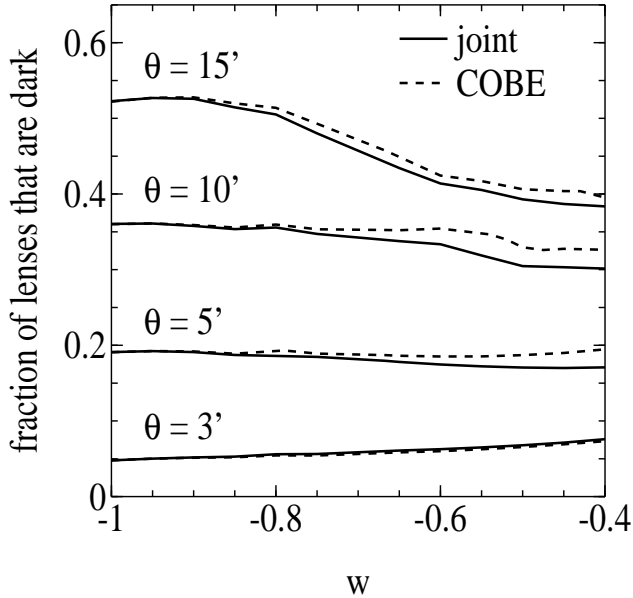,angle=0,width=3.4in}}
\vspace{1.2cm}
\caption{The fraction of lenses that are dark as a function of $w$ for aperture 
sizes $\theta=3$, $5$, $10$ and $15$ arcmin. Results are shown for both the COBE 
normalization with $\Omega_0=0.3$ (\emph{dashed} lines) and the joint COBE--cluster 
abundance normalization with $\Omega_0=\Omega_0(w)$  (\emph{solid} lines). As $\theta$ 
increases, the fraction of lenses that are dark rises significantly.}
\end{figure}
As mentioned above, the number-count ratio of dark to virialized lenses is an observable that is 
much less sensitive to observational noise than is the redshift distribution and number count 
of weak lenses. Unfortunately, for aperture sizes $\theta$ (defined in Section 2) less than 
10$'$ in radius the ratio is fairly constant over a broad range in $w$, as we show in 
Figure 11. The ratio varies more strongly if the aperture size is increased to 15$'$. 
In particular, for $\theta = 15'$ there is a $\sim 20$\% difference between the 
$\Lambda$CDM model and $w=-0.6$, so that differentiating the two models to a 3$\sigma$ 
significance requires the detection of $\sim 600$ virialized lenses or equivalently a sky 
coverage of $\sim 50$ degree$^2$. Although using the non-virialized lens fraction requires 
large survey coverage for modest constraints on $w$, its principal advantage (in addition to 
being relatively insensitive to observational noise) is that it is not very sensitive to the 
chosen method of normalization; for any aperture size both the joint normalization and the COBE 
normalization with fixed $\Omega_0$ yield similar dependences on $w$. Therefore, unlike the 
case for weak-lens sky-density or redshift distribution predictions, uncertainties in
$\sigma_8$ and $\Omega_0$ do not strongly 
affect the predicted ratio of dark to virialized lenses. Incidentally, although aperture 
sizes greater than $\sim 15'$ yield ratios with even stronger $w$ dependences, noise 
contributions from large-scale structure become significant at such large angular 
distances from the lens center (Hoesktra 2002). It is therefore not practical to make 
measurements at radii well beyond 15$'$. 

\begin{figure}
\vspace{-2.3cm}
\centerline{\hbox{\hspace{2.5cm}} \psfig{file=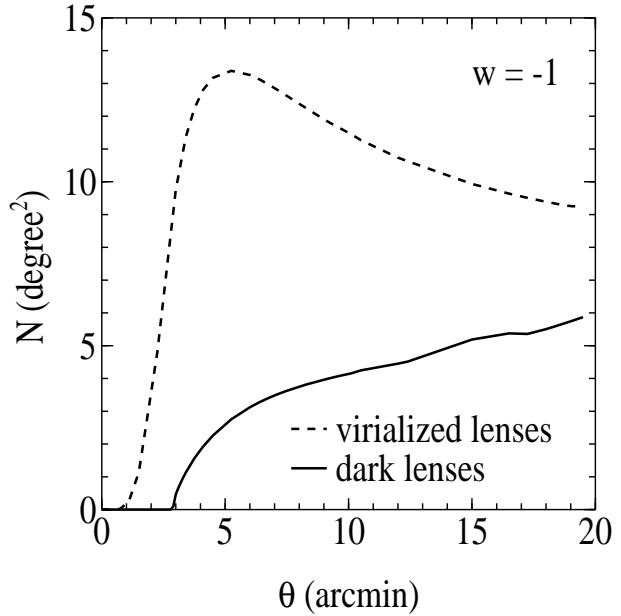,angle=0,width=3.4in}}
\vspace{1.2cm}
\caption{The total number of virialized lenses (\emph{dashed} line) and non-virialized 
lenses (\emph{solid} line) per square degree as a function of the aperture size 
$\theta$ for a $\Lambda$CDM cosmology. While the number count of virialized lenses 
peaks at $\theta=5$ arcmin and declines thereafter, the number count of 
non-virialized lenses increases almost linearly for $\theta > 5$ arcmin.}
\end{figure}
As an aside, while the ratio of dark to virialized lenses does not have a particularly strong $w$ 
dependence, it does have a strong $\theta$ dependence; only $\sim 5$\% of lenses are dark when 
$\theta= 3'$ but $\sim 50$\% when 
$\theta= 15'$. In Figure 12 we plot the number of virialized and dark lenses as a function of 
$\theta$ for the $\Lambda$CDM model. As $\theta$ increases from 3$'$ to 15$'$ the sky density of 
dark lenses increases from zero to 5 per square degree while the sky density of virialized lenses 
peaks at $\theta = 5'$ and gradually declines for larger aperture sizes. Figure 13 explains 
this trend. 
For an overdensity of mass $M = 5 \times 10^{14} M_{\odot}$ we plot, as a function of redshift, 
$\theta_{\rm{vir}}$, the projected angular size of the virialization radius, and 
$\theta_{\rm{max}}$, 
the projected angular size of the maximum radius that produces a detectable lens (i.e., 
$\theta_{\rm{max}}=R_{\rm{max}}(z)/D_{\rm{d}}(z)$ where 
$R_{\rm{max}}^3=3M / 4 \pi \Delta_{\rm{min}}(z)$). For $\theta_{\rm{max}} > \theta_{\rm{vir}}$ 
an overdensity can be non-virialized and still 
produce a detectable lensing signal (i.e., a dark lens). However, since $\theta$ defines the 
maximum observable angular scale, for sufficiently small $\theta$  
there is no range in redshift such that $\theta > \theta_{\rm{max}} > \theta_{\rm{vir}}$, in 
which case non-virialized overdensities cannot produce a detectable lens. In general we find 
that the minimum aperture size needed 
to detect dark lenses is $\sim 3'$. For larger $\theta$, the area below  $\theta_{\rm{max}}$ 
and above $\theta_{\rm{vir}}$ 
has a substantial relative increase while the area below $\theta_{\rm{vir}}$ has just a 
mild relative increase. After taking into account the fact that the aperture mass 
$M_{\rm{ap}} (\theta)$ decreases with increased $\theta$, this translates to an increase 
in the sky density of dark lenses and a decrease in the sky density of virialized lenses 
for $\theta > 5'$. The fraction of lenses that are dark therefore increases with aperture size.

\section{Discussion and Conclusions} 

We have examined the possibility of using the measured abundance of weak gravitational 
lenses to constrain a principal property of the dark energy, its equation-of-state parameter $w$. 
Since dark energy modifies both the background cosmology of the universe and the growth 
of structure it will necessarily have an effect on the efficiency of weak lensing. The 
goal of this paper was to determine the nature and strength of the effect. 

The change in the background cosmology with $w$ influences the predicted weak lens abundance 
in essentially three ways. First, the size of comoving volume elements shrink with increasing 
$w$. Second, the distance-redshift relation is modified, thereby shifting the location of the 
lensing-kernel maximum (i.e., where the combination of angular diameter distances 
$D_{\rm{ds}} D_{\rm{d}}/ D_{\rm{s}}$ peaks). Third, since the evolution of the background 
matter density is modified by the dark energy, the density of a given halo relative to the 
background density changes with $w$. This, in turn, affects the strength of a halo's lensing 
signal; the larger the overdensity the stronger the signal. While the volume term is 
explicitly factored into the expression for the weak-lens sky density [equation (22)], the 
latter two effects are incorporated into the signal-to-noise estimator for which we use the 
aperture-mass technique introduced by Schneider (1996).

The change in the growth of structure with $w$ is somewhat more subtle. The dark energy modifies 
both the rate of structure growth and the amplitude of the matter power spectrum. To determine 
the former we solved the spherical-collapse model with dark energy included. Though growth occurs 
more rapidly as $w$ increases, the overall effect on the $\Delta$ -- $\delta$ map, needed to 
relate the minimum overdensity required to produce a detectable lens, $\Delta_{\rm{min}}$, to a 
corresponding linear-theory overdensity $\delta_{\rm{min}}$, is fairly small. Similarly, the 
linear-theory overdensity at collapse $\delta_{\rm{c}}$ does not vary much with $w$. The effect 
on $\Delta_{\rm{vir}}$ is more significant, however. As $w$ increases, structures require 
substantially greater overdensities in order to reach virial equilibrium because they collapse
sooner, when the universe was younger and hotter. 

\begin{figure}
\vspace{-2.3cm}
\centerline{\hbox{\hspace{2.7cm}} \psfig{file=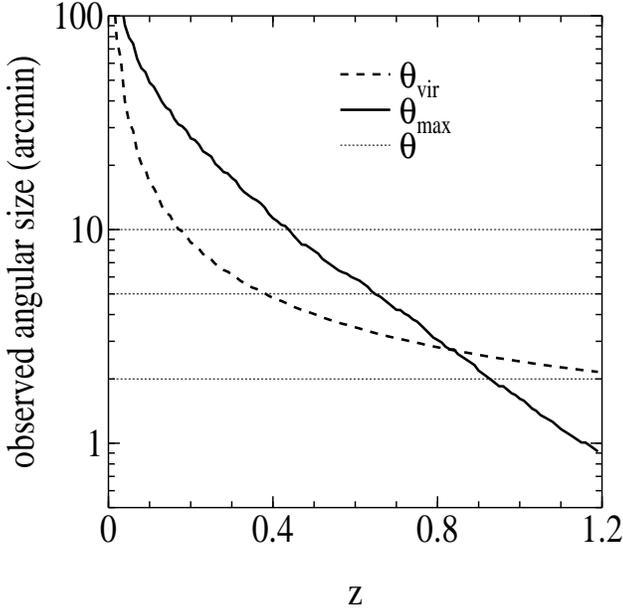,angle=0,width=3.4in}}
\vspace{1.2cm}
\caption{The observed angular size of $\theta_{\rm{vir}}$ (\emph{dashed} line) and 
$\theta_{\rm{max}}$ (\emph{solid} line) as 
a function of redshift for an overdensity of mass $M = 5 \times 10^{14} M_\odot$ for a 
$\Lambda$CDM cosmology. If $\theta_{\rm{max}} > \theta_{\rm{vir}}$ an overdensity need 
not be virialized to produce a detectable lensing signal. However, the range in redshift 
over which a non-virialized lens can be detected is limited by the aperture size 
$\theta$ (e.g., \emph{thin, dotted} lines), which defines the maximum observable angular scale. 
For $\theta \la 3$ arcmin virtually no dark lenses can be detected.}
\end{figure}
To determine how the power-spectrum amplitude, $\sigma_8$, varies with $w$ we considered three 
possible approaches. One was to normalize to the X-ray cluster abundance as was done in WS98. 
Another was to normalize to the COBE measurements of CMB anisotropies on large angular scales. 
These two 
approaches predict similar values of $\sigma_8$ for the $\Lambda$CDM model. However, if all 
cosmological parameters are held fixed as $w$ varies, the values of $\sigma_8$ are no longer in 
accordance. This is because the cluster abundance approach is accounting for the earlier forming, 
and hence hotter, galaxy clusters in models with $w > -1$. The COBE normalization, on the other 
hand, is accounting for the increase in the  Integrated Sachs-Wolfe (ISW) effect as $w$ increases 
(cf., BPB). Given these differing influences, the two approaches are not expected to 
yield the same $\sigma_8$ when all the cosmological parameters are held fixed to those of the 
$\Lambda$CDM model while $w$ is varied. This suggests a third approach to normalizing 
the power spectrum; namely, let the parameters vary with $w$ such that the cluster abundance 
normalization matches the COBE normalization. In practice we accomplished this by letting just 
$\Omega_0$ vary with $w$, as it is the parameter most degenerate with $\sigma_8$. The resulting 
range in $\Omega_0$ for $-1 < w \la -0.4$ was found to be $0.3 < \Omega_0 < 0.4$ and hence within 
observational uncertainties. Though all three normalization approaches predict that $\sigma_8$ 
decreases with $w$, the difference in the magnitude of the decrease between the approaches is 
significant. As a result, each predicts substantially different variations in the weak-lens 
abundance with $w$.    
 
Having determined all the dark energy effects, we computed the redshift distribution and sky 
density of weak lenses as a function of $w$. As in WK02, we distinguished between two classes 
of lenses, those that have collapsed and virialized and those that have not. This distinction 
is based on the expectation that the virialized lenses, being in a relaxed state, are X-ray 
and/or optically luminous. The non-virialized lenses, being as at earlier stage in the overdensity 
evolutionary cycle, are expected to be X-ray underluminous because the observed X-ray luminosity 
function has a steep dependence on the total virialized mass within a halo. Furthermore, though 
the typical mass of both lens types is $\sim few \times 10^{14} M_{\odot}$, the sky density of 
galaxies within the non-virialized lenses is expected to be smaller than in the virialized lenses 
because they have not yet collapsed and hence have larger radii (see WK02 for more details). 

We found that the variation in the redshift distribution and the sky density of both lens types
with $w$ depends strongly on the power-spectrum-normalization approach. If $\Omega_0$ is fixed 
and $\sigma_8$ is normalized to the COBE measurements, there is a significant variation in the 
abundances with $w$. In particular, the sky density of both virialized lenses and non-virialized 
lenses drops by a factor of two from $w = -1$ to $w = -2/3$. This decline, a result of the 
significant decrease in $\sigma_8$ with $w$, occurs despite the faster formation of structure 
for $w > -1$. If, on the other hand, $\Omega_0$ is 
allowed to vary with $w$ such that the COBE normalization matches the cluster-abundance 
normalization, the redshift distributions and sky density change very little with $w$; between 
$w =-1 $ and $w = -2/3$ the sky density of both lens types varies by just $\sim 20$\%. This 
insubstantial variation is the result of an increase in $\Omega_0$ with $w$ and a less 
significant drop in $\sigma_8$ with $w$ as compared to the COBE normalization with 
$\Omega_0$ fixed. Obtaining a strong constraint on $w$ from the sky density or redshift 
distribution of weak lenses therefore appears to be contingent on improved measurements of 
$\Omega_0$ from independent observations. 

Perhaps more promising is the possibility of utilizing the observed ratio of dark lenses to 
virialized lenses. Unlike measurements of the absolute sky density of weak lenses, the 
ratio is not very sensitive to the amount of observational noise in the weak-lensing maps since 
the abundance of both dark lenses and virialized lenses are equally affected by noise. Similarly, 
the ratio does not vary significantly over a wide range in cosmological parameters so that 
uncertainties due to the $\Omega_0-w$ degeneracy are minimized. We found that for aperture 
sizes of $\sim 15'$ the ratio varies by about 20\%, dropping from $0.5$ to $0.4$, between 
the $\Lambda$CDM model and $w = -0.6$. We also showed that the ratio of dark to virialized 
lenses increases with aperture size, in effect because larger apertures enable the 
detection of the more extended radii of the non-virialized lenses. 

Weak lensing has already been shown to be a powerful probe of the matter distribution in the 
universe (see e.g., Bartelmann \& Schneider 2001). It also has the potential to help constrain 
the amount and nature of the dark energy. Huterer (2002) showed that given reasonable prior
information on other cosmological parameters, the weak-lensing convergence power spectrum can 
impose constraints on the dark energy comparable to those of upcoming type Ia supernova and 
number-count surveys of galaxies and galaxy clusters. Constraining the dark energy 
from absolute measurements of weak-lens abundances will likely prove difficult, however. 
The variation in the weak lens sky density with $w$ is sufficiently small that modest 
uncertainties in $\Omega_0$ (and observational noise) can mask the effect of the dark energy.  
More auspicious is the possibility of utilizing the relative abundance of dark lenses to 
virialized lenses to constrain $w$. Future weak-lensing projects such as the the VISTA survey, 
the SNAP mission, and LSST (see Tyson et al. 2002 for a discussion of its great promise as a 
probe of dark energy) are expected to provide the wide-field surveys needed for this technique 
to be viable.

\section*{ACKNOWLEDGMENTS}
We thank R. Caldwell for helpful suggestions. NNW acknowledges the support of an NSF Graduate 
Fellowship. This work was supported by NSF AST-0096023, NASA NAG5-9821, and 
DoE DE-FG03-92-ER40701.

\label{lastpage}
\end{document}